\documentclass[12pt]{iopart}
\expandafter\let\csname equation*\endcsname\relax
\expandafter\let\csname endequation*\endcsname\relax
\usepackage{amsmath, amssymb}
\usepackage{graphicx}
\usepackage{dcolumn}
\usepackage{bm}

\usepackage[english]{babel}
\usepackage[utf8x]{inputenc}
\usepackage[T1]{fontenc}
\usepackage[colorinlistoftodos]{todonotes}
\usepackage{hyperref}
\usepackage{braket}
\usepackage{subcaption}
\DeclareMathAlphabet{\mathpzc}{OT1}{pzc}{m}{it}
\usepackage{mwe}
\usepackage{titlesec}
\usepackage{babel,blindtext}
\usepackage{mathrsfs}
\usepackage{gensymb}
\usepackage{sidecap}
\usepackage{float}
\DeclareMathOperator{\sech}{sech}
\usepackage{ctable}
\usepackage{appendix}
\usepackage{natbib}

\usepackage{subcaption}
\usepackage{ragged2e}
\DeclareCaptionJustification{justified}{\justifying}
\usepackage{pythontex}
\begin{document}

\title[]{Neutrino spin-flavor oscillations in solar environment}

\author{Sandeep Joshi \footnote{sjoshi@barc.gov.in} and Sudhir R. Jain \footnote{srjain@barc.gov.in}} 

\address{Nuclear Physics Division, Bhabha Atomic Research Centre, Mumbai 400085, India\\
Homi Bhabha National Institute, Anushakti Nagar, Mumbai 400094, India}

\begin{abstract}
 We study the phenomenon of neutrino spin-flavor oscillations due to solar magnetic fields. This allows us to examine how significantly the electron neutrinos produced in the solar interior undergo a resonant spin-flavor conversion. We construct analytical models for the solar magnetic field in all the three regions of the Sun. Neutrino spin-flavor oscillations in these magnetic fields are examined by studying the level crossing phenomenon and comparing the two cases of zero and non-zero vacuum mixing respectively. Results from the Borexino experiment are used to place an upper limit on the magnetic field in the solar core.  Related phenomena such as effects of matter on neutrino spin transitions and differences between Dirac and Majorana transitions in the solar magnetic fields are also discussed.      

\end{abstract}

\maketitle

   \maketitle

\section{Introduction}

The study of solar neutrinos and their oscillation phenomenology has revealed many facets of the physics of neutrinos. The Ray Davis experiment, which started in the $1960$'s in Homestake mine, was the first experiment to detect solar neutrinos reaching Earth. After several years of operation, the experiment reported that there is about a two-third deficit in the observed solar neutrino flux compared to the  standard solar model calculation (\citealt{Cleveland:1998nv}).  The deficit was further confirmed by other solar neutrino experiments, notably SAGE, GALLEX and Super-Kamiokande (SK) (\citealt{Abdurashitov:1994bc, Anselmann:1995ag, Hampel:1996qd, Fukuda:1996sz}). This discrepancy between the observed rate of neutrino flux and its theoretical prediction is called the solar neutrino problem. One of the ways to resolve the problem was suggested by Pontecorvo on the basis of mixing between different neutrino flavors. He showed that if neutrinos have a non-zero mass then the neutrino flavor mixing will give rise to oscillations among different neutrino flavors (\citealt{Bilenky:1978nj}). Thus electron neutrinos produced in the Sun may convert to some other flavor of neutrinos on their way to Earth and become undetectable. The problem was finally resolved when the Sudbury Neutrino Observatoryv(SNO) detected neutrinos from all three flavors in the solar neutrino flux, which proved that there must be transitions among the three active neutrino flavors (\citealt{Ahmad:2002jz}).  However, if vacuum neutrino oscillation alone were responsible for these flavor transitions, one would also be able to detect seasonal variation in the neutrino flux rate due to eccentricity of Earth's orbit. The $^{8}B$ neutrino spectrum in the SK experiment exhibited no such variation (\citealt{Hosaka:2005um}). The mechanism of flavor transitions that is most favored by data is the adiabatic resonant conversion due to neutrino-matter interactions, also known as the Mikheev-Smirnov-Wolfenstein (MSW) effect. Wolfenstein showed that the coherent forward scattering of neutrinos with electrons, protons and neutrons will induce an additional potential which will modify the effective mass and mixing of neutrinos in the medium (\citealt{Wolfenstein:1977ue}). In a medium with variable density, such as the Sun, these matter effects can lead to enhanced transitions between $\nu_e$ and $\nu_{\mu}/\nu_{\tau}$, even for small vacuum mixing angles (MSW-SMA) (\citealt{Mikheev:1986if, Mikheev:1986wj}). However, most of the solar neutrino data, including data from the KamLAND experiment and recent data from the Borexino experiment, have established the large mixing angle (MSW-LMA) solution to the solar neutrino problem  (\citealt{Abe:2008aa, Agostini:2018uly,  Robertson:2012ib,  Maltoni:2015kca, Wurm:2017cmm}).  

Another idea that was a popular candidate for the solution of the solar neutrino problem was spin precession of neutrinos in the magnetic field of the Sun. It was shown that if neutrinos have sufficiently large magnetic moment then the solar magnetic field can give rise to spin precession $\nu_{eL} \xrightarrow{} \nu_{eR}$, which will cause a deficit in the solar $\nu_e$ flux (\citealt{Cisneros:1970nq,Okun:1986na}). This solution was partly supported by data from  the Homestake experiment which observed anticorrelation between the neutrino flux and sunspot activity (\citealt{Davis:1994jw}). However, measurements from other experiments observed no such correlation (\citealt{Fukuda:1996sz}). Subsequently, the KamLAND experiment ruled out the spin-precession solution by placing a strong constraint on the flux of antineutrinos coming from the Sun (\citealt{Eguchi:2003gg}). A related effect due to neutrinos having non-zero transition magnetic moments is called resonant spin-flavor precession (RSFP) which results in both spin-flip and flavor change of neutrinos (\citealt{Lim:1987tk, Akhmedov:1988uk}). This effect arises due to the combination of matter and magnetic field on neutrino propagation and is similar to the MSW resonance, and can take place in transverse (\citealt{Akhmedov:1988uk}) as well as longitudinal magnetic fields (\citealt{Akhmedov:1988hd}). Also, the neutrino spin and spin-flavor transitions can give rise to other interesting quantum mechanical effects such as non-vanishing geometric phases (\citealt{Joshi:2016unj, Joshi:2017vpi}),  which demonstrate the intimate connection between the geometry of neutrino spin trajectory in the projective Hilbert space  and neutrino spin transition probabilities.  

Having determined the  basic oscillation parameters for solar neutrinos, the present effort is to search for sub-leading effects in the solar neutrino transitions which can give important clues for phenomena beyond the standard model. Various studies have been done to look for effects of non-standard interactions (NSI) (\citealt{Farzan:2017xzy}), dark matter imprints on the neutrino spectrum (\citealt{Lopes:2018wgp}), non-radiative neutrino decay (\citealt{Aharmim:2018fme}) and the combined effect of NSI and spin-flavor precession (SFP) (\citealt{Yilmaz:2016ilw}). In this paper, we study the possible sub-leading effects caused by spin-flavor transitions due to neutrino propagation in the solar magnetic field. 

The neutrino electromagnetic coupling is given by the Hamiltonian $H_{EM} = \frac{1}{2}  \bar{\nu}\mu \sigma_{\mu\nu} \nu F^{\mu \nu}$ + h.c., where $\mu$ is the neutrino magnetic moment matrix. For the case of Dirac neutrinos, the hermicity of the Hamiltonian requires $\mu^{\dagger} = \mu$. On the other hand, for Majorana neutrinos CPT symmetry requires the magnetic moment matrix to be anti-symmetric, which results in vanishing diagonal magnetic moments (\citealt{Schechter:1981hw}). This difference in the magnetic moment matrix gives rise to different spin-flavor transition probabilities for Dirac and Majorana neutrinos. The diagonal magnetic moment for a Dirac neutrino in the minimally extended standard model (MESM) including massive neutrinos is $\mu_\nu \approx 3 .2 \times 10^{-19} (m_\nu/1 \mbox{eV}) \mu_B$, where $m_\nu$ is the neutrino mass (\citealt{Marciano:1977wx,Lee:1977tib}). The  off-diagonal magnetic moments for both Dirac and Majorana neutrinos are further suppressed due to Glashow–Iliopoulos–Maiani (GIM) mechanism (\citealt{Pal:1981rm}).  However, the current best experimental bounds on the neutrino magnetic moment are in the range $\mu_{\nu} \leq (2-10)\times 10^{-11} \mu_B$   (\citealt{Giunti:2014ixa, Giunti:2015gga, Borexino:2017fbd}). Thus, the sensitivity of the present experiments is many orders away from the MESM predictions. To bridge this gap, many theoretical models have been postulated (see (\citealt{Giunti:2014ixa}) for detailed references)  which avoid the GIM suppression and predict neutrino magnetic moment in the range ($10^{-10}- 10^{-14})\mu_B$. For example, in the  left-right symmetric model, there is right-handed current that mixes with the Standard Model (SM)  left-handed current. In this model, one obtains $\mu_{\nu_e} = 2 \times 10^{-13} \mu_B \sin 2 \phi$, where $\phi$ is the mixing angle between the right and left-handed currents (\citealt{Shrock:1982sc, Fukugita:1987ti}). However, the mixing angle is expected to be small $\phi < 0.05$, thus limiting $\mu_{\nu_e}< 10^{-14} \mu_B$ (\citealt{Giunti:2014ixa, Fukugita:1987ti}). To obtain larger magnetic moments,  charged scalar particles are added to the left-right symmetric model (\citealt{Fukugita:1987ti, Fukugita:2003en, Babu:1987be}). The charged scalar contributions can give rise to magnetic moments in the range $\mu_\nu \sim (10^{-11}- 10^{-10}) \mu_B$.

In the present work, we examine the effects of magnetic moments $\sim 10^{-11} \mu_B$ on the solar neutrino transition probabilities for both the cases of Dirac and Majorana neutrinos. In particular, we first perform calculations for the approximate  case of vanishing vacuum mixing and show that the spin-flavor evolution equations can be reduced to a form which has an exact solution. We then study the actual case of non-zero mixing angle and the effects of the level crossing phenomenon on neutrino transition probabilities and use the results to place bounds on the solar magnetic fields. In the previous work along these lines by various authors (\citealt{TorrenteLujan:2003cx,Akhmedov:2002mf,Miranda:2004nz, Guzzo:2005rr,Chauhan:2005pn, Balantekin:2004tk, Friedland:2005xh, Das:2009kw}), several bounds have been obtained for both Dirac and Majorana spin-flavor transitions for different magnetic field configurations. 
 
The magnitude of the spin-flavor transitions depend mainly on the strength of the magnetic field at the location of the SFP resonance. This in turn depends on the detailed magnetic field profile of the Sun, which is not very well known, especially in the interior regions of the Sun. In Section 2, we discuss  current bounds on the solar magnetic field in various regions of the Sun and its effect on neutrino spin polarization. We also discuss the effective two-flavor model for neutrino spin-flavor precession. In Section 3, we  show that in the approximate case of vanishing mixing angle the resulting set of equations posses analytically exact solutions. We also derive bounds on the solar magnetic fields using the existing experimental results. We then examine the effect of non-zero vacuum mixing on neutrino transition probabilities. Finally we discuss the results in the last Section.

\section{Neutrino spin precession in solar magnetic fields}
The magnetic field in different regions of the Sun exhibits different characteristic behaviors (\citealt{Friedland:2005xh}). In the solar convective zone (CZ) the magnetic fields are believed to be generated from a dynamo mechanism active at its base. The current data from helioseismology points to a thin shear layer at the bottom of the CZ, known as a tachocline, which generates a large scale toroidal magnetic field. The strength of the magnetic field is predicted to be in the range 10-100 kG (\citealt{fan2009magnetic}).  On the other hand, the radiative zone (RZ) magnetic field may have its origin in the formation of the Sun. Once formed, this primordial field might have been frozen in the RZ and the solar core without protruding much into the CZ (\citealt{dicke1982magnetic}). The bound on the large scale toroidal magnetic field in the RZ ranges from 5-7 MG (\citealt{Friedland:2002is}) to 30 MG (\citealt{Couvidat:2002bs}).  For the solar core, magnetic field bounds vary widely from 30 G (\citealt{boruta1996solar}) to 7 MG (\citealt{Antia:2002kk}).

\begin{figure} 
\Centering \includegraphics[width=100mm]{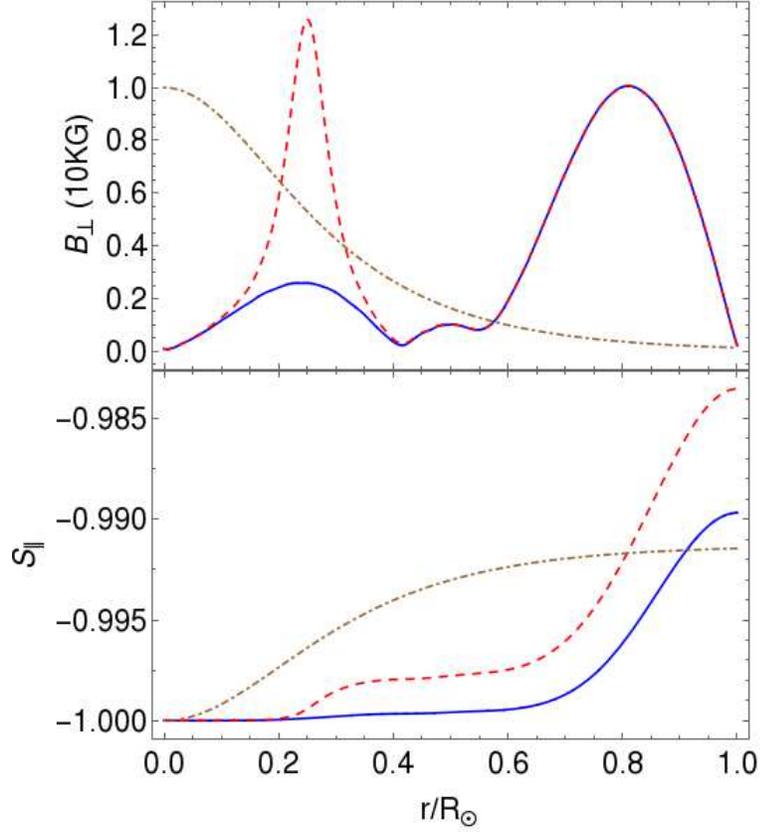}
\caption{\label{fig1} The longitudinal neutrino spin polarization $S_\parallel$ as it propagates in the magnetic field of the Sun. The solid curve is the magnetic field obtained by solving solar MHD equations in \citealt{Miranda:2000bi}. The dashed curve is given by equation \eqref{brz} and the dot-dashed curve by equation \eqref{bcore}. The peak magnetic field for both models is taken to be $\approx 10^4$ G. }
\label{fig:polarization}
\end{figure}

Based on the above bounds, we choose two profiles to simulate the magnetic field in the Sun.  In the first model we implement the field profile given by \citealt{Miranda:2000bi} and add an RZ magnetic field 
\begin{equation} \label{brz}
    B_{\perp RZ} (r) = B_0 \sech [34.75(r/R_\odot- 0.25)].
\end{equation}
The profile is chosen such that $B_{RZ}$ in the CZ is
negligible compared to the CZ magnetic field and also becomes very small near the solar core. For the second model, we select a field profile which peaks in the solar core and is expressed as
\begin{equation}\label{bcore}
    B_{\perp} (r)= B_0 \sech 5 r/R_\odot.
\end{equation}
First we consider the neutrino spin precession as it propagates in the solar magnetic field neglecting the effect of matter and flavor mixing. The change in neutrino spin polarization in this case is described by the equation 

\begin{equation}
    \frac{d \bm{S}}{dr} = 2 \mu_\nu \bm{S} \times \bm{B}_{\perp}(r) ,
\end{equation}
where for $B_{\perp}$ we apply the two magnetic field profiles in \eqref{brz}, \eqref{bcore} and $\mu_{\nu} \approx 10^{-11} \mu_B$. As can be seen in Figure \ref{fig:polarization}, the change in neutrino spin polarization can be sufficient even with peak fields $\sim 10^4$. The change in helicity of solar neutrinos can also affect the $\nu - e$ scattering (\citealt{Barranco:2017zeq}). 

Now if we include the matter potential term $V$ which affects left and right helicity states differently, then the neutrino propagation can be described by a Schr\"{o}dinger-like equation (\citealt{Giunti:2014ixa})
\begin{equation}
    i \frac{d}{dr}\begin{pmatrix}
    \nu_L \\ \nu_R
    \end{pmatrix} = \begin{pmatrix}
    V(x) & \mu_\nu B_{\perp} \\
    \mu_\nu B_{\perp} &  0
    \end{pmatrix}
    \begin{pmatrix}
    \nu_L \\ \nu_R
    \end{pmatrix}.
\end{equation}
 For the case of constant V and $B_{\perp}$, the change in neutrino helicity is expressed as 
 \begin{equation}\label{prob-1}
     P_{\nu_L \rightarrow \nu_R}(x) = \frac{(2 \mu_\nu B_{\perp})^2}{V^2+(2 \mu_\nu B_{\perp})^2} \sin ^2 \bigg(\frac{1}{2} \sqrt{V^2+(2 \mu_\nu B_{\perp})^2} x \bigg).
 \end{equation}
 Thus, matter potential is expected to further suppress the change in  neutrino helicity in solar magnetic fields. 
 
 Now considering two neutrino flavors, we finally include the effects of neutrino masses and mixing angle $\theta_{12}$. In this case, the effective Hamiltonian becomes a $4 \times 4$ matrix. For the case of Dirac neutrinos, the effective Hamiltonian in the $(\nu_{eL}, \nu_{\mu L}, \nu_{eR}, \nu_{\mu R})^T$ basis is given by \citealt{Giunti:2014ixa}
 \begin{equation} \label{hamil:dirac}
     H_D = \begin{pmatrix}
    -\frac{ \Delta m^2}{4 E}\cos{2 \theta_{12}}+ V_e &  \frac{ \Delta m^2}{4 E} \sin{2 \theta_{12}} & \mu_{ee} B_{\perp} & \mu_{e \mu} B_{\perp} \\
     \frac{ \Delta m^2}{4 E} \sin{2\theta_{12}} &  \frac{ \Delta m^2}{4 E}\cos{\theta_{12}}+ V_{\mu} & \mu_{\mu e} B_{\perp} & \mu_{\mu \mu} B_{\perp} \\
     \mu_{ee} B_{\perp} & \mu_{\mu e} B_{\perp} &  -\frac{ \Delta m^2}{4 E} \cos{2 \theta_{12}} &  \frac{\Delta m^2}{4 E} \sin{2 \theta_{12}} \\
     \mu_{e \mu} B_{\perp} & \mu_{\mu \mu} B_{\perp} &  \frac{ \Delta m^2}{4 E} \sin{2 \theta_{12}} &  \frac{ \Delta m^2}{4 E} \cos{2 \theta_{12}}
    \end{pmatrix},
 \end{equation}
 where $V_e = \sqrt{2} G_F (n_e - n_n/2) $ and $V_\mu = - G_F n_n /\sqrt{2}$ are matter potentials for left handed electron and muon neutrinos respectively, $n_e$ and $n_n$ denote the number densities of electrons and neutrons respectively and $\Delta m^2 = \Delta m_{21}^2$ is the neutrino mass-squared difference. For the Majorana case the vanishing diagonal terms $\mu_{ee}$ and $\mu_{\mu \mu}$ result in the following Hamiltonian in the $(\nu_{eL},  \nu_{\mu L}, \bar{\nu}_e, \bar{\nu}_\mu)^T$ basis (\citealt{Giunti:2014ixa})
 
 \begin{equation} \label{hamil-maj}
     H_M= \begin{pmatrix}
    -\frac{\Delta m^2}{4 E}\cos{2 \theta_{12}}+ V_e &  \frac{ \Delta m^2}{4 E} \sin{2 \theta_{12}} & 0 & \mu_{e \mu} B_{\perp} \\
     \frac{ \Delta m^2}{4 E} \sin{2\theta_{12}} &  \frac{ \Delta m^2}{4 E}\cos{\theta_{12}}+ V_{\mu} & -\mu_{\mu e} B_{\perp} & 0 \\
     0 & -\mu_{\mu e} B_{\perp} &  -\frac{ \Delta m^2}{4 E} \cos{2 \theta_{12}}- V_e &  \frac{\Delta m^2}{4 E} \sin{2 \theta_{12}} \\
     \mu_{e \mu} B_{\perp} & 0 &  \frac{ \Delta m^2}{4 E} \sin{2 \theta_{12}} &  \frac{ \Delta m^2}{4 E} \cos{2 \theta_{12}}- V_{\mu}
    \end{pmatrix}.
 \end{equation}
 
 Suppression due to the potential term in the two component case in Eq. \eqref{prob-1} can now be lifted due to resonant transitions. The electron neutrinos produced in the Sun can undergo multiple resonances in the presence of a magnetic field. The usual MSW resonance $\nu_{eL} \leftrightarrow \nu_{\mu L}$ takes place at the location  $x_{MSW}$
 \begin{equation} \label{loc-msw}
         \frac{\rho(x) Y_e}{m_n}\Bigg |_{x= x_{MSW}} = \frac{\Delta m^2 \cos{2 \theta_{12}}}{2 \sqrt{2} G_F E}.
    \end{equation}
In addition, there is spin-flavor resonance $\nu_{eL} \leftrightarrow \nu_{\mu R}$ which always occurs before the MSW resonance. The location of the spin-flavor resonance is given by
\begin{equation} \label{loc-sfp}
    \frac{\rho(x) Y_e^{\mbox{eff}}}{m_n}\Bigg |_{x= x_{SFP}}= \frac{\Delta m^2 \cos{2 \theta_{12}}}{2 \sqrt{2} G_F E},
\end{equation}
where $\rho(x)$ is matter density inside the Sun, $m_n$ is the neutron mass,  $Y_e$ is the electron fraction  and 
\begin{equation} \label{yeff}
Y_e^{\rm eff}= \begin{cases}
(3 Y_e-1)/2 \quad \mbox{for}\quad \nu{_e{_L}}\leftrightarrow \nu_{\mu{_R}},\\
(2 Y_e-1) \quad \mbox{for} \quad  \nu_{e{_L}}\leftrightarrow \bar{\nu}_{\mu}.
\end{cases}
\end{equation}

\begin{table}[h!] 
\centering
\begin{tabular}{|c c c |} 
 \hline
 E (MeV) & $\nu{_e{_L}}\leftrightarrow \nu_{\mu{_R}}$ & $\nu_{e{_L}}\leftrightarrow \bar{\nu}_{\mu}$ \\
 \hline 
 2.5 & 0.057 & 0.027  \\ 
 5.0 & 0.156 & 0.142 \\ 
 10.0 & 0.230 & 0.218 \\
 15.0 & 0.268 & 0.257 \\[1ex]
 \hline
 \end{tabular}
 \caption{ The location of SFP resonance in the Sun (in units $r/R_\odot$) for different neutrino energies.}
 \label{table-1}
\end{table}

The location of resonance for different neutrino energies are provided in  \autoref{table-1} using the electron density profile from the solar model BS2005 of Bahcall, Serenelli and Basu (\citealt{Bahcall:2004pz}).  We have used $\Delta m^2 = 7.4 \times 10^{-5}$ eV$^2$ and $\theta_{12} = 33.8 ^{\degree}$ throughout the paper. For neutrinos with energy below $2$ MeV, the resonant density required is too high to occur in the Sun. Thus only the high energy $^{8}B$ neutrinos are expected to be affected by these effects.

The solutions of the neutrino evolution equation with spin-flavor Hamiltonian \eqref{hamil:dirac} and \eqref{hamil-maj} are difficult to solve for arbitrary varying density and magnetic fields. However, analytical (\citealt{Aneziris:1991qj}) and semi-analytic (\citealt{Yilmaz:2017igr}) solutions exist for different cases. In the next section, we will study the case of zero vacuum mixing which gives rise to equations admitting exact analytical solutions.

\section{An analytical model for zero vacuum mixing} \label{sec-3}
\begin{figure}[h!]
\Centering \includegraphics[width= 100mm]{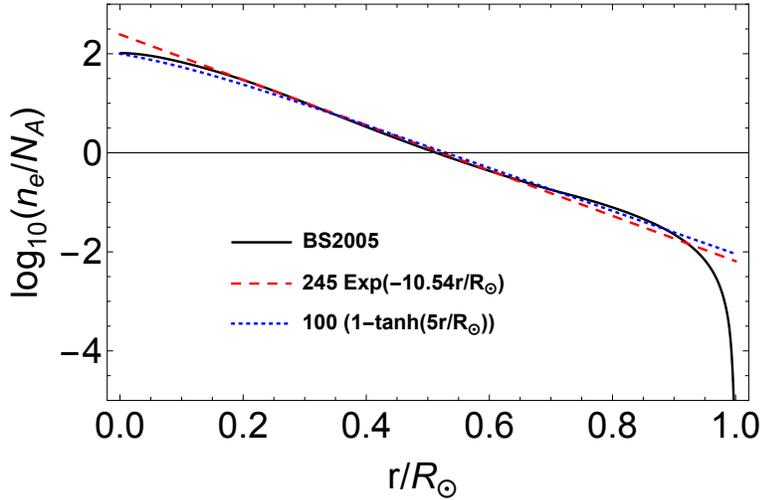}
\caption{\label{fig1} Electron number density variation vs. radial distance in the Sun. The \textit{solid line} represents the solar model BS(2005) and the \textit{dashed curves} are analytical approximations.}
 \label{fig:density}
\end{figure}

For the case of $\theta_{12} = 0$, only the SFP resonance can contribute to the neutrino transitions. In this case the effective Hamiltonian becomes a $2 \times 2 $ matrix in the channel $\nu_{eL} \leftrightarrow \nu_{\mu R}/\bar{\nu}_{\mu}$ :
\begin{equation} \label{hamiltonian-1}
    H = \begin{pmatrix}
    \frac{-\Delta m^2 }{4E}+ \frac{\delta V}{2} & \mu_{e \mu} B \\
    \mu_{e \mu} B & \frac{\Delta m^2}{4E} - \frac{\delta V}{2}
    \end{pmatrix},
\end{equation}
where $\delta V = \sqrt{2} G_F \rho Y_e^{\rm eff}/m_N$, with $Y_e^{\rm eff}$ defined by Eq. \eqref{yeff}. As can be seen from Eq. \eqref{hamiltonian-1}, the main input required to study spin-flavor transitions is the profile of number density of electrons and neutrons, and the magnetic field along the neutrino trajectory. The electron number density in the solar model BS(2005) is shown in Figure \ref{fig:density}. However, for obtaining numerical solutions various approximations, are applied (\citealt{Pal:1991pm}). Here we use the approximation 
\begin{equation}\label{density}
    n_e(r) = 100 [1- \tanh(5 r /R_\odot)] N_A \quad cm^{-3},
\end{equation}
where $N_A$ is the Avogadro's number, which gives a reasonably good approximation apart from the region near the surface of the Sun. 

\begin{figure}[t]
\Centering \includegraphics[width= 120mm]{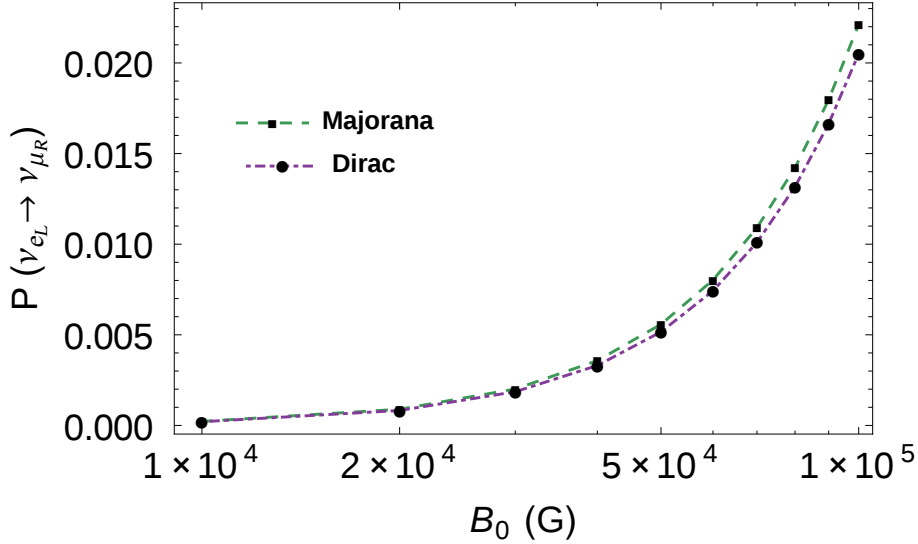}
\caption{Transition probability of Dirac and Majorana neutrinos obtained from the solution of equation \eqref{ode-1}. Here the neutrinos are assumed to be produced at the center of the Sun with energy $E= 10$ MeV. }
\label{fig:dk-prob1}
\end{figure}

Now the equation for the neutrino flavor $\nu_{eL}$ with Hamiltonian \eqref{hamiltonian-1} becomes a second order  ordinary differential equation given by 
\begin{equation}\label{ode-1}
        \frac{d^2\nu_{eL}}{dt^2}- \Bigg(\frac{\mu \dot{B}}{\mu B}+ i \xi\Bigg)\frac{d\nu_{eL}}{dt}+ \Bigg(\phi^2+ i \frac{d\phi}{dt}+ (\mu B)^2- i \phi \frac{\mu \dot{B}}{\mu B}+ \phi \xi\Bigg)\nu_{eL} = 0,
    \end{equation}
where we have defined 
\begin{align}
    \phi=& -\frac{\Delta m^2}{4E} + \frac{1}{\sqrt{2}} G_F n_e,\\
    \xi=& \begin{cases}
- \frac{1}{\sqrt{2}}G_F n_n \quad \mbox{for}\quad \nu{_e{_L}}\rightarrow \nu_{\mu{_R}},\\
- \sqrt{2}G_F n_n \quad \mbox{for} \quad  \nu_{e{_L}}\rightarrow \bar{\nu}_{\mu}.
\end{cases}
\end{align}
In general, it is possible to solve this equation numerically to obtain the survival probability of electron neutrinos. However for the case when magnetic field is given by Eq. \eqref{bcore} and density is expressed by Eq. \eqref{density}, the set of equations reduces to the well known Demkov-Kunike model, which has exact solutions (\citealt{suominen1992population, kenmoe2016demkov}). The analytical solution  is provided by Eq. \eqref{nuel} and can be used to calculate the neutrino transition probability $P(\nu_{eL} \rightarrow \bar{\nu}_{\mu R}; R_\odot)$. The resulting solution plotted in Figure \ref{fig:dk-prob1} depicts the difference for the two cases of Dirac and Majorana neutrinos. For sufficiently low magnetic fields, the difference in the transition probability of the two cases is not significant. However, for large magnetic field there can be a detectable difference in the Dirac and Majorana neutrinos. 
 
If we assume that inside the Sun the transitions are driven dominantly by SFP resonance, and that outside the Sun  the transitions are mainly due to the large vacuum mixing angle, then the probability for the electron neutrinos produced inside the Sun to reach the Earth's surface as electron antineutrinos is given by (\citealt{Akhmedov:2002mf})
\begin{equation} \label{prob-2}
\begin{aligned}[b]
    P(\nu_e \rightarrow \bar{\nu}_e )=& P(\nu_{eL} \rightarrow \bar{\nu}_{\mu R}; R_\odot) P( \bar{\nu}_{\mu R}\rightarrow  \bar{\nu}_{e R}; R_{es}) \\
    =&  P(\nu_{eL} \rightarrow \bar{\nu}_{\mu R}; R_\odot) \sin^2 2 \theta_{12} \sin^2 \big(\frac{\Delta m^2 R_{es}}{4 E}\big),
 \end{aligned} 
\end{equation}
where $R_{es}$ is the average distance between Earth and Sun.

For the above model, the result from the Borexino experiment can be used to obtain bounds on the maximum magnetic field $B_0$ at the center of the Sun. The Borexino experiment gives an upper limit on the neutrino transition probability for $^{8}B$ neutrinos $P_{\nu_{e}\rightarrow \bar{\nu}_e} < 1.3 \times 10^{-4}$ at $90 \%$ C.L. for $E_{\bar{\nu}} > 1.8$ MeV (\citealt{Bellini:2010gn}). 

\vspace{0.1cm}
 
Now the transition probability $P(\nu_{eL} \rightarrow \bar{\nu}_{\mu R}; R_\odot)$ in Eq. \eqref{prob-2} is obtained from Eq. \eqref{nuel} by averaging over the $^{8}B$ neutrino production region in the Sun (\citealt{Bahcall:2004pz}). Using this, we  calculate the the mean probability in the energy region ($2< E < 15$) MeV with $1$ MeV/bin. For Majorana neutrinos we obtain $\langle P \rangle =1.18 \times 10^{-4}$ for $B_0 = 3 \times 10^4$ G and $\langle P \rangle = 2.1 \times 10^{-4}$ for $B_0 = 4 \times 10^4$ G. Whereas for the case of Dirac neutrinos we obtain $\langle P \rangle =1.0 \times 10^{-4}$ for $B_0 = 3 \times 10^4$ G and $\langle P \rangle = 1.8 \times 10^{-4}$ for $B_0 = 4 \times 10^4$ G. Thus the consistency with the Borexino result requires $B_0 \leq 3 \times 10^{4}$ G in both cases.
 Hence, this analysis presents us a useful bound on the magnetic field in the solar core. This bound lies in between the various other bounds discussed in the previous section. However, this limiting case obtained by substituting $\theta_{12} = 0$ inside the solar region over estimates the transition probability by pushing the SFP resonance deeper into the solar interior where the strength of the magnetic field is higher. Thus we expect the actual bound on the magnetic field to be higher in the full treatment with all the flavors taken into consideration.

For the case when magnetic field is given by Eq.\eqref{brz} in the RZ of the Sun, such analytical solutions of Eq. \eqref{ode-1} are not possible. In this case, since the magnetic field is significantly weaker at the SFP location, we do not expect significant transitions. Hence, the bounds on the RZ magnetic field will be comparatively weaker.    
 
\section{Including effects of $\theta_{12}$}
Adding the effects of the vacuum mixing leads to  the full Hamiltonian \eqref{hamil:dirac} and \eqref{hamil-maj} for Dirac and Majorana neutrinos respectively. However, since there is no resonant production of  $\nu_{e R}/ \bar{\nu}_e$, we set its amplitude to zero which yields the effective $3 \times 3$ Hamiltonian for the Majorana neutrinos
\begin{equation} \label{hamil2-maj}
     H_M= \begin{pmatrix}
    \frac{-\Delta m^2}{4 E}\cos{2 \theta_{12}}+ V_e &  \frac{ \Delta m^2}{4 E} \sin{2 \theta_{12}} & \mu B_{\perp} \\
     \frac{ \Delta m^2}{4 E} \sin{2\theta_{12}} &  \frac{ \Delta m^2}{4 E}\cos{\theta_{12}}+ V_{\mu}  & 0 \\
     \mu B_{\perp} & 0 &   \frac{ \Delta m^2}{4 E} \cos{2 \theta_{12}}- V_{\mu}
    \end{pmatrix},
 \end{equation}
 and a similar one for the Dirac neutrinos. In this case, we have two resonances described by equations \eqref{loc-msw} and \eqref{loc-sfp}. However at the location of both  resonances, the Hamiltonian is dominated by large off-diagonal term $\Delta m^2 \sin 2\theta_{12}/4E$ . Thus merely fulfilling the SFP resonant condition in equation \eqref{loc-sfp} is not sufficient to drive large transitions due to the magnetic field. In this case, it is more appropriate to go to mass eigenbasis where such large vacuum mixing terms are absent (\citealt{Friedland:2005xh}). The Hamiltonian in the mass eigenbasis can be obtained by performing a rotation on the flavor eigenstates 
 \begin{equation}
     H_M \rightarrow R_{12}^\dagger H_M R_{12},
 \end{equation}
and diagonalizing the resultant matrix, where $R_{12}$ is the rotation matrix in the $(12)$ plane. We obtain 
\begin{equation} \label{hamil3-maj}
    H_M^D = \begin{pmatrix}
    \Delta_D & 0 & \mu B \cos \theta_D \\
    0  & - \Delta_D & \mu B \sin \theta_D \\
    \mu B \cos \theta_D & \mu B \sin \theta_D & - \kappa_M
    \end{pmatrix},
\end{equation}
where \begin{align} 
    \Delta_D =& \sqrt{\bigg(-\frac{\Delta m^2}{4E} \cos 2 \theta_{12}+ \frac{1}{\sqrt{2}} G_F n_e\bigg)^2+\bigg(\frac{\Delta m^2}{4E} \sin 2 \theta_{12}\bigg)^2}, \\
    \theta_D=& - \frac{1}{2} \tan^{-1}\Bigg( \frac{\frac{\Delta m^2}{4E} \sin 2 \theta_{12}}{-\frac{\Delta m^2}{4E} \cos 2 \theta_{12}+ \frac{1}{\sqrt{2}} G_F n_e}\Bigg), \\
    \kappa_M = & -\frac{\Delta m^2}{4E} \cos 2 \theta_{12}+ \frac{1}{\sqrt{2}} G_F (n_e- 2 n_n).
\end{align}
 
\begin{figure}
\centering
\begin{subfigure}[b]{.65\textwidth}
  \centering
  \includegraphics[width=0.9\linewidth]{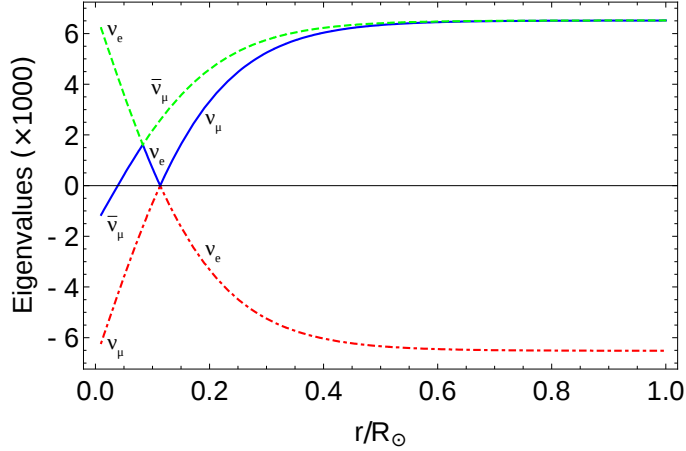}
  \caption{\centering}
  \label{fig:sub1}
\end{subfigure}%
\\
\vspace{0.5cm}
\begin{subfigure}[b]{.65\textwidth}
  \centering
  \includegraphics[width=0.9\linewidth]{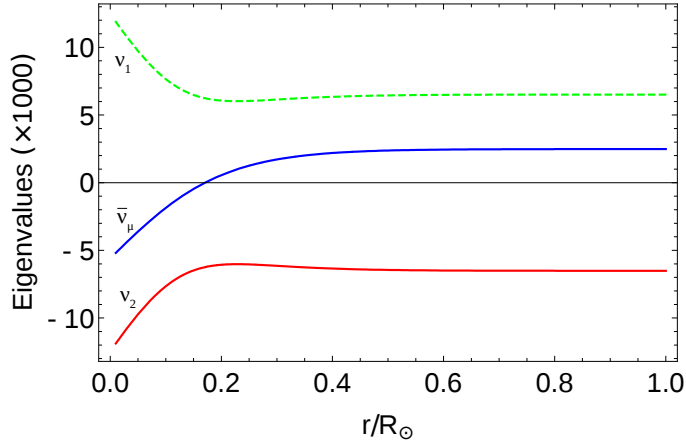}
  \caption{\centering}
  \label{fig:sub2}
\end{subfigure}
\caption{Eigenvalues of the Hamiltonian for $E= 10$ MeV neutrinos: (a) in the flavor basis, equation \eqref{hamil2-maj} for $\theta_{12}\approx 0 $. The two level \textit{crossing points} correspond to SFP and MSW resonances. (b) in the mass eigenbasis, equation \eqref{hamil3-maj} for $\theta_{12}= 33.8^{\degree}$. The \textit{dashed/dot-dashed lines} correspond to $\nu_1/\nu_2$ respectively and the solid line represents $\bar{\nu}_\mu$. Here we have used $B_0= 10^6$ G and the eigenvalues are in dimensionless units. } 
\label{fig:eigenvalues}  
\end{figure}
 
 In Figure \ref{fig:eigenvalues} we plot the eigenvalues of the Majorana Hamiltonian equations \eqref{hamil2-maj} and \eqref{hamil3-maj} in flavor and mass basis respectively. In the flavor basis. depicted in Figure \ref{fig:sub1}, one can see the level crossing at two different points. The lower one corresponds to SFP resonance while the higher one is the MSW resonance. The electron neutrinos are produced predominantly in the heavier mass eigenstate (dashed curve in Figure \ref{fig:sub1}). At the SFP crossing point, the transition between the neutrino states $\nu_e \leftrightarrow \bar{\nu}_\mu$ is driven by the strength of the magnetic field at the location of the level crossing. Assuming the level crossing to be adiabatic, the $\nu_e$ eigenstate is now represented by the solid curve in Figure \ref{fig:sub1} while the dashed curve corresponds now to $\bar{\nu}_\mu$. The electron neutrino then goes through another resonance at the MSW crossing point. After this second level crossing, the $\nu_e$ state now corresponds to the dot-dashed curve which is the lower mass eigenstate while $\nu_\mu$ is the upper mass eigenstate (solid curve).
 
 However, this notion of resonant flavor conversion is valid only for small mixing angles (\citealt{Friedland:2000rn}). For large values of mixing angle, the  mass eigenbasis describes the situation more accurately. Comparing Figure \ref{fig:sub1} and \ref{fig:sub2}, it is seen that the level crossing which was present for the case $\theta_{12}\approx 0$ is  now absent. Again, if the electron neutrinos are produced in the heavier mass eigenstate (dashed curve in Figure \ref{fig:sub2}), they now will not encounter any level crossing resonance such as those in Figure \ref{fig:sub1}. Thus merely fulfilling the resonant conditions in Eqs. \eqref{loc-msw} and \eqref{loc-sfp} is not sufficient for resonant conversion and these conditions are valid only for small mixing angle. A general condition for resonant conversion can also be derived which holds for both small and large mixing angles (\citealt{Friedland:2005xh}). 
 
\begin{figure}[t] 
\Centering \includegraphics[width= 100mm]{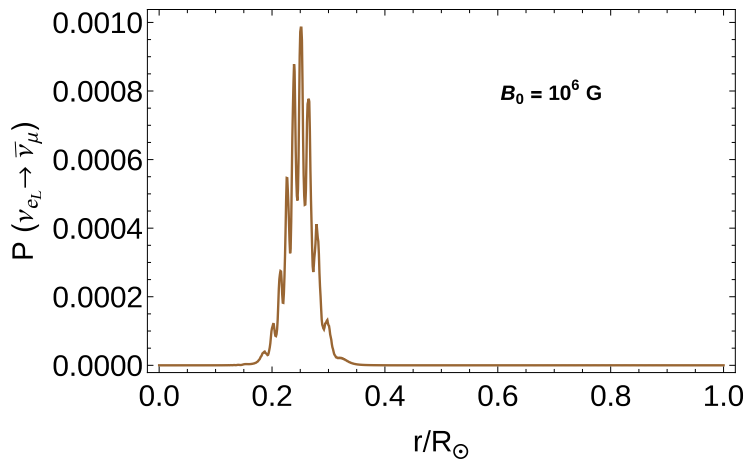}
\caption{ The variation of probability   $P(\nu_{eL} \rightarrow \bar{\nu} _{\mu R})$ with distance inside the Sun for maximum RZ magnetic field $B_0 = 10^6$ G. The neutrinos are assumed to be produced at the center of the Sun and $E= 10$ MeV.   } 
\label{fig:3-flav}
\end{figure}

 An examination of the neutrino transitions as it propagates in the Sun reveals further details about the neutrino evolution in this general case. Working with Hamiltonian \eqref{hamil3-maj} we can see at the point of neutrino production near the solar core the diagonal terms are $\Delta_D \sim 4 \times 10^{-12} $ eV for $E=10$ MeV, while the magnetic field term $\mu B \sim 6 \times 10^{-16} $ eV for $B \sim 10^4$ G. Thus there is a difference of about four orders of magnitude and the transitions will be absent.  As the neutrino propagates to the lower density regions in the RZ, the eigenlevels come closer.  At $r \approx 0.2 R_\odot$ we have $\Delta_D \sim 2 \times 10^{-12}$ eV while the magnetic field now increases to about $10^6$G, thus $\mu B \sim 6 \times 10^{-14}$eV. There is still a difference of about an order of magnitude, however now there can be small $\nu_{eL} \leftrightarrow \bar{\nu}_\mu$ transitions driven by the magnetic field as can be acertained in Figure \ref{fig:3-flav}. These conversions persist as long as the ratio $\Delta_D/\mu B \sim 0.1$. However beyond $r= 0.4 R_\odot$, the magnetic field gradually falls off to values $< 10^{5}$ G (see Figure \ref{fig:polarization}), and the corresponding  transitions also die out.  Thus after the partial conversion  of the neutrinos $\nu_e \rightarrow \bar{\nu}_\mu$ in the region $r \approx (0.2- 0.4) R_{\odot}$, the neutrino reverts back to being predominantly in the eigenstate $\nu_1$. As the neutrinos propagate towards the CZ, they will again encounter an increasing magnetic field. However due to the strong bounds on the magnetic field in this region having peak field $B_0 < 10^5$ G, the diagonal splitting  terms $\Delta_D >> \mu B$ and there will be no significant transitions due to magnetic fields. Thus assuming the neutrinos are produced in the eigenstate $\nu_1$ in the Sun, they will exit the Sun in the same eigenstate and buried magnetic field in the RZ having strength $\sim 10^6$ G is not sufficient to cause any appreciable level crossing. Thus the transitions are suppressed to a great extent. 
 
We can write the neutrino transition probability 
\begin{equation} \label{prob-3}
    P(\nu_{eL} \rightarrow \bar{\nu} _{\mu R}) = \sum_i P(\nu_{eL} \rightarrow \nu_{i}) P(\nu_{i}\rightarrow \nu_{\mu R}),
\end{equation}
where $P(\nu_{eL} \rightarrow \nu_{i})$ is the probability that the electron neutrino is produced in mass eigenstate $\nu_{i}$ and $P(\nu_{i}\rightarrow \nu_{\mu R})$ is the probability of transition $\nu_i \rightarrow \bar{\nu}_{\mu R}$ under the effect of magnetic field. Since the Hamiltonian in equation \eqref{hamil3-maj} for the Majorana neutrinos can be effectively decoupled into two $2 \times 2$ blocks, we can write 
\begin{equation} \label{prob-4}
  P(\nu_{eL} \rightarrow \bar{\nu} _{\mu R}) = \cos^2 \theta_D(r_i) P(\nu_1 \rightarrow \bar{\nu}_{\mu R}) + \sin^2 \theta_D(r_i) P(\nu_2 \rightarrow \bar{\nu}_{\mu R}) ,
\end{equation}
where $\theta_D(r_i)$ is the mixing angle at the neutrino production point $r_i$. The probabilities $P(\nu_{i}\rightarrow \nu_{\mu R})$ can be evaluated numerically to give the total transition probability in equation \eqref{prob-4}. 

\section{Comparison with Borexino results }
\begin{figure}[t] 
\Centering \includegraphics[width=120mm]{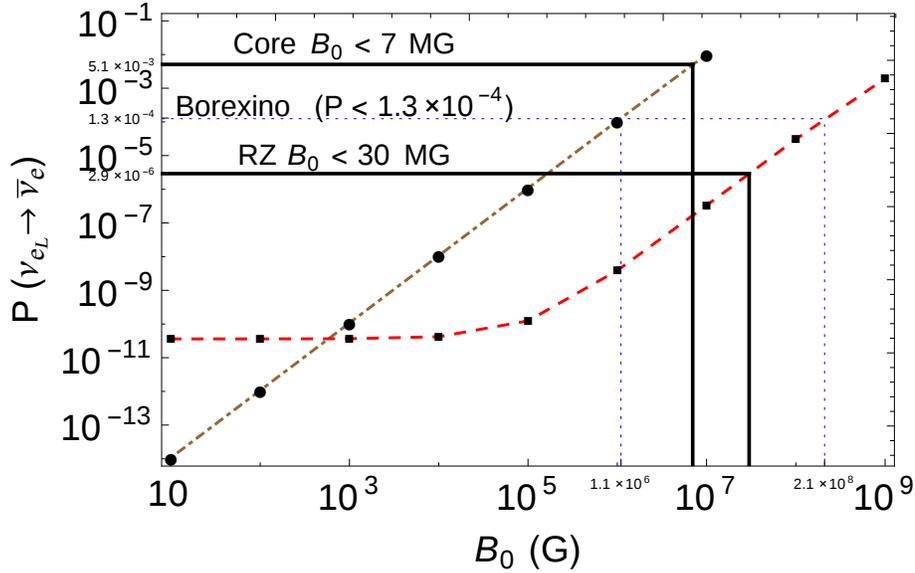}
\caption{  The probability of solar electron neutrino ($E= 10$ MeV) to anti-neutrino conversion at the Earth's surface (Eq.\eqref{prob-2}) and comparison with Borexino results. The \textit{dashed(red) curve and dotdashed(brown) curve} show the probability $P(\nu_e\rightarrow\bar{\nu}_e)$ calculated using the two field profiles marked with respective curves in Figure \ref{fig:polarization}. The \textit{dotted(blue) line} signifies that the current upper bound   $ P(\nu_e \rightarrow \bar{\nu}_e) < 1.3 \times 10^{-4}$ from the Borexino experiment corresponds to a bound of $2.1\times10^8$ G on the RZ magnetic field and to a bound of $1.1\times10^6$ G on the core magnetic field. The \textit{solid(black) lines} mark the helioseismological bounds of $30$ MG and $7$ MG on the RZ and solar core magnetic fields respectively.} 
\label{fig:3-flav-borexino}
\end{figure} 
The most stringent constraints on the  anti-neutrino flux are given by the Borexino experiment (\citealt{Bellini:2010gn}), which reported an upper limit of  $ \phi_{\bar{\nu}_e}  < 760 $ cm$^{-2}$ s$^{-1}$ on the $^{8}B$ flux. For an undistorted $^{8}B$ neutrino spectrum, the solar anti-neutrino flux at the surface of Earth is given by 
\begin{equation}
    \phi_{\bar{\nu}_e} = \phi_{\nu_e}(^{8}B) P(\nu_e \rightarrow \bar{\nu}_e),
\end{equation}
where the value of total $^{8}B$ neutrino flux is $\phi_{\nu_e}(^{8}B) = 5.88 \times 10^6$ cm$^{-2}$ s$^{-1}$ (\citealt{Bellini:2010gn}). Thus Borexino placed an upper bound of $ P(\nu_e \rightarrow \bar{\nu}_e) < 1.3 \times 10^{-4}$.
 
The solar electron neutrino transition probability $P(\nu_e \rightarrow \bar{\nu}_e)$ at the Earth's surface can be calculated using equation \eqref{prob-2}, where $P(\nu_{eL} \rightarrow \bar{\nu}_{\mu R}; R_\odot)$ is numerically evaluated using equation \eqref{prob-4} and is averaged over the $^{8}B$ neutrino production region in the Sun (\citealt{Bahcall:2004pz}). To put appropriate bounds on the solar magnetic field, we plot in Figure \ref{fig:3-flav-borexino} the probability $P(\nu_e \rightarrow \bar{\nu}_e)$ against the peak magnetic field for the case of Majorana Hamiltonian \eqref{hamil2-maj}. The two curves in Figure \ref{fig:3-flav-borexino} correspond to the two magnetic field profiles shown in  Figure \ref{fig:polarization}, one peaking at the center of the Sun and other in the RZ, in accordance with the existing helioseismological bounds. In Figure \ref{fig:3-flav-borexino}, we also show that the Borexino limit (\citealt{Bellini:2010gn}) intersects the two curves at points corresponding to the maximum allowed peak magnetic field. For the first case when the magnetic field peaks in the RZ, using the Borexino limit we obtain the value of peak magnetic field $B_0< 2.1\times 10^8$G. Thus the Borexino data is unable to constrain the existing bound of $B_0<30 $MG in the solar RZ, which corresponds to the probability  $ P(\nu_e \rightarrow \bar{\nu}_e) <  2.9\times 10^{-6}$ and hence to an upper limit  $ \phi_{\bar{\nu}_e}  < 17 $ cm$^{-2}$ s$^{-1}$ of the anti-neutrino flux. This requires an improvement by almost two orders of magnitude in the sensitivity of $\bar{\nu}_e$ detection. However, the same analysis with magnetic field peaking in the solar core provides very useful bounds which constrain some of the existing solar models. The Borexino limit in this case yields an upper bound of $B_0< 1.1 \times 10^6$ G, which is almost a factor of one-seventh of the current largest bound on the core magnetic field (\citealt{Antia:2002kk}). It is useful to compare this result with that obtained in Section \ref{sec-3}, where we obtained much stronger bound of $B_0< 8 \times 10^4$ G. This demonstrates that the two component approximation used frequently (e.g. (\citealt{Cuesta:2008te}) does not give the correct transition probability and it is more appropriate to take into account all possible channels in which the initially produced neutrino state may undergo resonant conversion.

Since the Borexino experiment continues to take data, it is natural to assume that future results will be able to place more stringent limits on the anti-neutrino flux. This in turn will be useful for placing stricter upper bounds on the solar magnetic field, especially in the solar core region where current helioseismological bounds vary widely in predictions.

\section{Conclusions}
In this paper, we have studied the phenomenon of neutrino spin-flavor oscillations in the Sun for neutrinos having sufficiently large magnetic moments $\sim 10^{-11} \mu_B$. We have constructed two models for solar magnetic field based on the current bounds on the magnetic field in different regions of the Sun. In the first model, one can have large magnetic field in the solar core and it tapers off with distance from the center. In the second model, we have a large magnetic field in the RZ which becomes negligible in the core region and in addition there is a CZ magnetic field, calculated in  \citealt{Miranda:2000bi}. It was shown that even a magnetic field $\sim 10^4$ G is sufficient to change the neutrino helicity as it comes out of the Sun. We have also obtained a novel parametrization for the electron density profile in the Sun, which provides a better approximation compared to the usual exponential parametrization. 

For the case of zero vacuum mixing and large magnetic field in the solar core, we obtain analytically exact solutions. This allows us to put strong bounds on the magnetic field in the solar core using results from the Borexino experiment. Also, the difference between the Dirac and Majorana neutrinos is significant only for magnetic fields $\sim 10^5$ G or more. We then examined the effects for the realistic case of large vacuum mixing angle and found that it has an effect in suppressing the $\nu_e \rightarrow \bar{\nu}_\mu$ transitions. The energy level diagrams distinctly demonstrate the difference between the two cases. Whereas in the case of small mixing angle we get enhanced transitions due to adiabatic level crossings. For the latter case of large vacuum mixing, the eigenstates of the Hamiltonian in the mass eigenbasis do not exhibit such crossing phenomenon. Thus the dominant terms are the diagonal terms and small transitions take place only in the RZ where the ratio of the two terms is $\sim 0.1$. Furthermore, the CZ fields do not affect the neutrino transitions.  The Borexino results are then utilized to place appropriate bounds on the two models of solar magnetic field. It is found that whereas the Borexino bounds are too weak to place any upper limit on the RZ magnetic field, for the solar core magnetic field we are able to place an upper bound $B_0< 1.1 \times 10^6$G. This is significant improvement over the existing bounds coming from helioseismology results. 

Based on the above results it can be seen that while the sub-leading effects on solar neutrinos due to spin-flavor transitions are likely to be very small for $\mu_\nu \sim 10^{-11} \mu_B$, with improved sensitivity, the future experiments will be able to place even stronger constraints on the neutrino magnetic moment as well as solar magnetic field. Thus the phenomenon of spin-flavor oscillations gives important information about the solar interior independent of helioseismological observations.  

\section{Acknowledgement}
The authors would like to acknowledge helpful comments and suggestions from the referee. 
\appendix
\section{Neutrino evolution equations and Demkov-Kunike model}
For the case when magnetic field and density of the Sun are given by Eqs. \eqref{bcore} and \eqref{density} the Hamiltonian \eqref{hamiltonian-1} can be written as 
\begin{equation} \label{hamiltonian-dk}
    H = \begin{pmatrix}
   \frac{-\Delta m^2 }{4E} + \frac{V_0}{2}(1- \tanh (5 r/R_{\odot}))   & \mu B_0 \sech(5r/R_{\odot}) \\
     \mu B_0 \sech(5r/R_{\odot}) & \frac{\Delta m^2 }{4E} - \frac{V_0}{2}(1- \tanh (5 r/R_{\odot})) 
    \end{pmatrix},
\end{equation}
where $V_0= \sqrt{2} G_F Y_e^{\rm eff}\rho_0/m_N$ with $\rho_0$ being the density at the solar center. We define
\begin{align}
    a=& -\frac{\Delta m^2 }{4E}+ \frac{V_0}{2},\\
    b=& -\frac{V_0}{2},\\
    c=& \mu B_{0}.
\end{align}
For ultra-relativistic neutrinos propagating along the radial direction in the Sun, the flavor equation \eqref{ode-1} can now be written as 
\begin{equation}
\begin{aligned}[b]\label{ode-2} 
    \frac{d^2 \nu_{eL}}{dr^2} + \frac{5}{R_\odot}\tanh(5r/R_\odot) \frac{d\nu_{eL}}{dr} +& \Big(c^2 \sech^2(5 r/R_\odot)+ (a+ b\tanh(5r/R_\odot))^2  \\& + \frac{5i}{R_\odot}(a\tanh(5r/R_\odot)+b)\Big)\nu_{eL}=0.
\end{aligned}
\end{equation}
Now substituting $z= (1+ \tanh(5r/R_\odot))/2$, Eq. \eqref{ode-2} becomes
\begin{equation}
\begin{aligned}[b]\label{ode-3}
&z(1-z) \frac{d^2 \nu_{eL}}{dz^2}+ \frac{1}{2}(1-2z)\frac{d\nu_{eL}}{dz}+ c^2\Big(\frac{ R_\odot}{5}\Big)^2 q(z) \nu_{eL}=0 ,   
\end{aligned}
\end{equation}
where
\begin{equation}
\begin{aligned}[b]
q(z)=& 1+ \frac{1}{4 c^2 z (1-z)}\Big(\big(a+ b(2 z-1)\big)^2 + \frac{5 i}{R_\odot}(a(2z-1)+b)\Big). 
\end{aligned}
\end{equation}
Finally the substitution $\nu_{eL} = z^{\mu} (1-z)^{\nu} u(z)$, where 
\begin{align}
    \mu =& -i (a-b)R_\odot/10, \\
    \nu =& i (a+b)R_\odot/10,
\end{align}
converts Eq. \eqref{ode-3} to a Gauss hypergeometric equation
\begin{equation}\label{ode-4}
    \begin{aligned}[b]
    z(1-z) \frac{d^2 u}{dz^2}+ (\gamma- (\alpha+ \beta+ 1)z)\frac{du}{dz}- \alpha \beta u(z) = 0,
    \end{aligned}
\end{equation}
where 
\begin{align}
    \alpha= & \frac{R_\odot}{10}\Big(ib + \sqrt{- b^2 + 4 c^2}\Big),\\
    \beta= &  \frac{R_\odot}{10}\Big(ib - \sqrt{- b^2 + 4 c^2}\Big),\\
    \gamma=& \frac{1}{2}- i(a-b) \frac{R_\odot}{5}.
\end{align}
Eq. \eqref{ode-4} has two linearly independent solutions which can be taken as (\citealt{Notzold:1987cq})
\begin{equation}
    \nu_{eL \pm} = z^{\pm \mu}(1-z)^\nu u_{\pm}(z),
\end{equation}
where $u_{\pm}(z)= u(z)|_{\mu \rightarrow \pm \mu}$
If the neutrinos are produced at the location $r_0$ inside the Sun, then the evolution of the state $\nu_{eL}$ is given by 
\begin{equation}\label{nuel}
    \begin{aligned}[b]
    \nu_{eL}(r)=& \cos^2 \theta_m e^{i \omega r_0} z^\mu(1-z)^\nu \hspace{0.1mm} _{2}F_{1}(\alpha, \beta, \gamma; z) \\&+ \sin^2 \theta_m e^{-i \omega r_0} z^{-\mu}(1-z)^\nu \hspace{0.1mm} _{2}F_{1}(\alpha, \beta, \gamma; z)|_{\mu \rightarrow - \mu},
    \end{aligned}
\end{equation}
where $\theta_m= \tan^{-1}(c/a)/2$, $\omega= \sqrt{(a)^2+(c)^2}$ and $_{2}F_{1}(\alpha, \beta, \gamma; z)$ is the  Gauss hypergeometric function. Since $b^2>> 4 c^2$, we can use $\alpha \approx \mu+ \nu, \beta \approx 0 $ and $\gamma = (1/2)+ 2 \mu$ for evaluating the survival probability given by $P_{ee}(r_0,r)= |\nu_{eL}(r)|^2 $. The transition probability $1- P_{ee}(r_0, r)$ is then averaged over the $^{8}B$ neutrino production region to put appropriate bounds on the magnetic field.

\bibliographystyle{raa.bst}
\newcommand{\newblock}{}
\bibliography{main.bbl}

\begin{thebibliography}{66}
\providecommand\natexlab[1]{#1}
\providecommand\JournalTitle[1]{#1}

\bibitem[Abdurashitov {et~al.}(1994)]{Abdurashitov:1994bc}
Abdurashitov, J.~N., {et~al.} 1994, Phys. Lett., B328, 234

\bibitem[Abe {et~al.}(2008)]{Abe:2008aa}
Abe, S., {et~al.} 2008, Phys. Rev. Lett., 100, 221803

\bibitem[Agostini {et~al.}(2017)]{Borexino:2017fbd}
Agostini, M., {et~al.} 2017, Phys. Rev., D96, 091103

\bibitem[Agostini {et~al.}(2018)]{Agostini:2018uly}
Agostini, M., {et~al.} 2018, Nature, 562, 505

\bibitem[Aharmim {et~al.}(2019)]{Aharmim:2018fme}
Aharmim, B., {et~al.} 2019, Phys. Rev., D99, 032013

\bibitem[Ahmad {et~al.}(2002)]{Ahmad:2002jz}
Ahmad, Q.~R., {et~al.} 2002, Phys. Rev. Lett., 89, 011301

\bibitem[Akhmedov(1988)]{Akhmedov:1988uk}
Akhmedov, E.~K. 1988, Phys. Lett., B213, 64

\bibitem[Akhmedov \& Khlopov(1988)]{Akhmedov:1988hd}
Akhmedov, E.~K., \& Khlopov, M.~{\relax Yu}. 1988, Mod. Phys. Lett., A3, 451

\bibitem[Akhmedov \& Pulido(2003)]{Akhmedov:2002mf}
Akhmedov, E.~K., \& Pulido, J. 2003, Phys. Lett., B553, 7

\bibitem[Aneziris \& Schechter(1992)]{Aneziris:1991qj}
Aneziris, C., \& Schechter, J. 1992, Phys. Rev., D45, 1053

\bibitem[Anselmann {et~al.}(1995)]{Anselmann:1995ag}
Anselmann, P., {et~al.} 1995, Phys. Lett., B357, 237, [Erratum: Phys.
  Lett.B361,235(1995)]

\bibitem[Antia(2002)]{Antia:2002kk}
Antia, H.~M. 2002, ESA Spec. Publ., 505, 71

\bibitem[Babu \& Mathur(1987)]{Babu:1987be}
Babu, K.~S., \& Mathur, V.~S. 1987, Phys. Lett., B196, 218

\bibitem[Bahcall {et~al.}(2005)]{Bahcall:2004pz}
Bahcall, J.~N., Serenelli, A.~M., \& Basu, S. 2005, Astrophys. J., 621, L85

\bibitem[Balantekin \& Volpe(2005)]{Balantekin:2004tk}
Balantekin, A.~B., \& Volpe, C. 2005, Phys. Rev., D72, 033008

\bibitem[Barranco {et~al.}(2017)]{Barranco:2017zeq}
Barranco, J., Delepine, D., Napsuciale, M., \& Yebra, A. 2017, arXiv:1704.01549

\bibitem[Bellini {et~al.}(2011)]{Bellini:2010gn}
Bellini, G., {et~al.} 2011, Phys. Lett., B696, 191

\bibitem[Bilenky \& Pontecorvo(1978)]{Bilenky:1978nj}
Bilenky, S.~M., \& Pontecorvo, B. 1978, Phys. Rept., 41, 225

\bibitem[Boruta(1996)]{boruta1996solar}
Boruta, N. 1996, The Astrophysical Journal, 458, 832

\bibitem[Chauhan {et~al.}(2005)]{Chauhan:2005pn}
Chauhan, B.~C., Pulido, J., \& Raghavan, R.~S. 2005, JHEP, 07, 051

\bibitem[Cisneros(1971)]{Cisneros:1970nq}
Cisneros, A. 1971, Astrophys. Space Sci., 10, 87

\bibitem[Cleveland {et~al.}(1998)]{Cleveland:1998nv}
Cleveland, B.~T., Daily, T., Davis, Jr., R., {et~al.} 1998, Astrophys. J., 496,
  505

\bibitem[Couvidat {et~al.}(2003)]{Couvidat:2002bs}
Couvidat, S., Turck-Chieze, S., \& Kosovichev, A.~G. 2003, Astrophys. J., 599,
  1434

\bibitem[Das {et~al.}(2009)]{Das:2009kw}
Das, C.~R., Pulido, J., \& Picariello, M. 2009, Phys. Rev., D79, 073010

\bibitem[Davis(1994)]{Davis:1994jw}
Davis, R. 1994, Prog. Part. Nucl. Phys., 32, 13

\bibitem[Dicke(1982)]{dicke1982magnetic}
Dicke, R. 1982, Solar Physics, 78, 3

\bibitem[Eguchi {et~al.}(2004)]{Eguchi:2003gg}
Eguchi, K., {et~al.} 2004, Phys. Rev. Lett., 92, 071301

\bibitem[Fan(2009)]{fan2009magnetic}
Fan, Y. 2009, Living Reviews in Solar Physics, 6, 4

\bibitem[Farzan \& Tortola(2018)]{Farzan:2017xzy}
Farzan, Y., \& Tortola, M. 2018, Front.in Phys., 6, 10

\bibitem[Friedland(2001)]{Friedland:2000rn}
Friedland, A. 2001, Phys. Rev., D64, 013008

\bibitem[Friedland(2005)]{Friedland:2005xh}
Friedland, A. 2005, arXiv:hep-ph/0505165

\bibitem[Friedland \& Gruzinov(2004)]{Friedland:2002is}
Friedland, A., \& Gruzinov, A. 2004, Astrophys. J., 601, 570

\bibitem[Fukuda {et~al.}(1996)]{Fukuda:1996sz}
Fukuda, Y., {et~al.} 1996, Phys. Rev. Lett., 77, 1683

\bibitem[Fukugita \& Yanagida(1987)]{Fukugita:1987ti}
Fukugita, M., \& Yanagida, T. 1987, Phys. Rev. Lett., 58, 1807

\bibitem[Fukugita \& Yanagida(2003)]{Fukugita:2003en}
Fukugita, M., \& Yanagida, T. 2003, {Physics of neutrinos and applications to
  astrophysics}

\bibitem[Giunti {et~al.}(2016)]{Giunti:2015gga}
Giunti, C., Kouzakov, K.~A., Li, Y.-F., {et~al.} 2016, Annalen Phys., 528, 198

\bibitem[Giunti \& Studenikin(2015)]{Giunti:2014ixa}
Giunti, C., \& Studenikin, A. 2015, Rev. Mod. Phys., 87, 531

\bibitem[Guzzo {et~al.}(2005)]{Guzzo:2005rr}
Guzzo, M.~M., de~Holanda, P.~C., \& Peres, O. L.~G. 2005, Phys. Rev., D72,
  073004

\bibitem[Hampel {et~al.}(1996)]{Hampel:1996qd}
Hampel, W., {et~al.} 1996, Phys. Lett., B388, 384

\bibitem[Haxton {et~al.}(2013)]{Robertson:2012ib}
Haxton, W.~C., Hamish~Robertson, R.~G., \& Serenelli, A.~M. 2013, Ann. Rev.
  Astron. Astrophys., 51, 21

\bibitem[Hosaka {et~al.}(2006)]{Hosaka:2005um}
Hosaka, J., {et~al.} 2006, Phys. Rev., D73, 112001

\bibitem[Joshi \& Jain(2016)]{Joshi:2016unj}
Joshi, S., \& Jain, S.~R. 2016, Phys. Lett., B754, 135

\bibitem[Joshi \& Jain(2017)]{Joshi:2017vpi}
Joshi, S., \& Jain, S.~R. 2017, Phys. Rev., D96, 096004

\bibitem[Kenmoe {et~al.}(2016)]{kenmoe2016demkov}
Kenmoe, M., Tchapda, A., \& Fai, L. 2016, J. Math. Phys., 57, 122106

\bibitem[Lee \& Shrock(1977)]{Lee:1977tib}
Lee, B.~W., \& Shrock, R.~E. 1977, Phys. Rev., D16, 1444

\bibitem[Lim \& Marciano(1988)]{Lim:1987tk}
Lim, C.-S., \& Marciano, W.~J. 1988, Phys. Rev., D37, 1368

\bibitem[Lopes \& Silk(2019)]{Lopes:2018wgp}
Lopes, I., \& Silk, J. 2019, Phys. Rev., D99, 023008

\bibitem[Maltoni \& Smirnov(2016)]{Maltoni:2015kca}
Maltoni, M., \& Smirnov, A.~{\relax Yu}. 2016, Eur. Phys. J., A52, 87

\bibitem[Marciano \& Sanda(1977)]{Marciano:1977wx}
Marciano, W.~J., \& Sanda, A.~I. 1977, Phys. Lett., B67, 303

\bibitem[Mikheev \& Smirnov(1986{\natexlab{a}})]{Mikheev:1986if}
Mikheev, S.~P., \& Smirnov, A.~{\relax Yu}. 1986{\natexlab{a}}, Sov. Phys.
  JETP, 64, 4, [Zh. Eksp. Teor. Fiz.91,7(1986)]

\bibitem[Mikheev \& Smirnov(1986{\natexlab{b}})]{Mikheev:1986wj}
Mikheev, S.~P., \& Smirnov, A.~{\relax Yu}. 1986{\natexlab{b}}, Nuovo Cim., C9,
  17

\bibitem[Miranda {et~al.}(2001)]{Miranda:2000bi}
Miranda, O.~G., Pena-Garay, C., Rashba, T.~I., Semikoz, V.~B., \& Valle, J.
  W.~F. 2001, Nucl. Phys., B595, 360

\bibitem[Miranda {et~al.}(2004)]{Miranda:2004nz}
Miranda, O.~G., Rashba, T.~I., Rez, A.~I., \& Valle, J. W.~F. 2004, Phys. Rev.,
  D70, 113002

\bibitem[Mosquera~Cuesta \& Lambiase(2008)]{Cuesta:2008te}
Mosquera~Cuesta, H.~J., \& Lambiase, G. 2008, Astrophys. J., 689, 371

\bibitem[Notzold(1987)]{Notzold:1987cq}
Notzold, D. 1987, Phys. Rev., D36, 1625

\bibitem[Okun {et~al.}(1986)]{Okun:1986na}
Okun, L.~B., Voloshin, M.~B., \& Vysotsky, M.~I. 1986, Sov. Phys. JETP, 64,
  446, [Zh. Eksp. Teor. Fiz.91,754(1986)]

\bibitem[Pal(1992)]{Pal:1991pm}
Pal, P.~B. 1992, Int. J. Mod. Phys., A7, 5387

\bibitem[Pal \& Wolfenstein(1982)]{Pal:1981rm}
Pal, P.~B., \& Wolfenstein, L. 1982, Phys. Rev., D25, 766

\bibitem[Schechter \& Valle(1981)]{Schechter:1981hw}
Schechter, J., \& Valle, J. W.~F. 1981, Phys. Rev., D24, 1883, [Erratum: Phys.
  Rev.D25,283(1982)]

\bibitem[Shrock(1982)]{Shrock:1982sc}
Shrock, R.~E. 1982, Nucl. Phys., B206, 359

\bibitem[Suominen \& Garraway(1992)]{suominen1992population}
Suominen, K.-A., \& Garraway, B. 1992, Phys. Rev., A45, 374

\bibitem[Torrente-Lujan(2003)]{TorrenteLujan:2003cx}
Torrente-Lujan, E. 2003, JHEP, 04, 054

\bibitem[Wolfenstein(1978)]{Wolfenstein:1977ue}
Wolfenstein, L. 1978, Phys. Rev., D17, 2369

\bibitem[Wurm(2017)]{Wurm:2017cmm}
Wurm, M. 2017, Phys. Rept., 685, 1

\bibitem[Yilmaz(2016)]{Yilmaz:2016ilw}
Yilmaz, D. 2016, Adv. High Energy Phys., 2016, 1435191

\bibitem[Yilmaz(2018)]{Yilmaz:2017igr}
Yilmaz, D. 2018, Turk. J. Phys., 42, 600

\end{thebibliography}

\end{document}